\documentclass[aps,twocolumn,floatfix,amsfonts,amssymb,notitlepage]{revtex4-1}
\usepackage{graphicx}
\usepackage{amsmath}
\usepackage{epstopdf}
\usepackage{amsbsy}
\usepackage{color}
\usepackage{dcolumn}
\usepackage{bm}
\usepackage{mathtools} 

\DeclareMathOperator{\sgn}{sgn}

\begin{document}

\title{Phonon mediated superconductivity in low carrier-density systems}
\author{Maria N. Gastiasoro, Andrey V. Chubukov, and Rafael M. Fernandes}
\affiliation{School of Physics and Astronomy, University of Minnesota, Minneapolis, MN 55455, USA.}

\date{\today}

\begin{abstract}
Motivated by the observation of superconductivity in SrTiO$_3$ and Bi,
we analyze phonon-mediated superconductivity in three-dimensional systems at low carrier density, when the
chemical potential $\mu$ (equal to Fermi energy at $T=0$) is comparable to or even smaller
 than the characteristic phonon frequency $\omega_L$. We consider the attractive part of the Bardeen-Pines pairing interaction, in which the frequency-dependent electron-phonon interaction is dressed by the Coulomb potential. This dressing endows the pairing interaction with momentum  dependence.
 We argue that the conventional Migdal-Eliashberg (ME) approximation becomes invalid
  when $\mu \leq \omega_L$ chiefly because
the dominant contribution
 to pairing comes from electronic states away from the Fermi surface.
  We obtain the pairing onset temperature, which is equal to $T_c$ in the absence of phase fluctuations, as a function of $\mu/\omega_L$.
     We find both analytically and numerically that $T_c$ increases as the ratio  $\mu/\omega_L$ becomes smaller.
     In particular, in the dilute regime, $\mu \rightarrow 0$, it holds that
  $T_c\propto\omega_L\left(\frac{Ry}{\omega_L}\right)^\eta$, where $\text{Ry}$ is the Rydberg constant and $\eta \sim 0.2$.
\end{abstract}
\maketitle

\section{Introduction}

 Studies of  superconductivity in systems with small Fermi surfaces (FSs)  attracted a lot of attention over the last decade. The most known and most studied systems of this kind are multi-band quasi-2D Fe-based superconductors with small hole and electron pockets~\cite{review_pnictides_1,*review_pnictides_2,*review_pnictides_3,*review_pnictides_4}.
 But superconductivity in 3D systems with small FSs and only one type of carriers also attracted a lot of attention in the past~\cite{Gurevich1962,Schooley64,Schooley65,Chernik1981,Takada1980,Ikeda1992,Grimaldi1995}, and the interest to this issue resurfaced in the last few years
   \cite{Lin14,Nakajima2015,Prakashaaf8227,Edge2015,Ruhman2016,Gorkov2016,Gorkov2017,Ruhman2016,Ruhman2017,
   Trevisan2018,Savary2017,Rowley2018,Lonzarich2018,Woelfle2018,Sadovskii2018,Sadovskii2018b}.
  The systems of  current interest for superconductivity at low electron density  include doped SrTiO$_3$, in which it has been long established that  superconductivity is still present at
 carrier densities as low as $n\sim 10^{18}$ cm$^{-3}$~\cite{Schooley64,Schooley65,Lin14}, and other materials
 such as Pb$_{1-x}$Tl$_x$Te~\cite{Chernik1981}, half-Heusler compounds~\cite{Nakajima2015}, and single crystal Bi~\cite{Prakashaaf8227}.
  The observations of superconductivity in these dilute systems motivate the general theoretical investigation of how superconductivity emerges in a system with a small FS and, consequently, small chemical potential $\mu$, counted from the bottom of the band.

The pairing mechanisms in dilute 3D  systems
 are widely debated, and several pairing scenarios have been suggested, particularly  for  strontium titanate, where superconductivity emerges in the vicinity of a ferroelectric transition~\cite{Takada1980,Edge2015,Ruhman2016,Gorkov2017}.
 We will not discuss material-specific  mechanisms here and instead focus on several general aspects of s-wave pairing due to an exchange of a longitudinal optical phonon with a frequency $\omega_L$.

 For systems  with sizable electron density, where
 $\mu$ well exceeds $\omega_L$, the analysis of phonon-mediated s-wave pairing is  traditionally done using Migdal-Eliashberg (ME) formalism
  ~\cite{Eliashberg,Migdal,Scalapino1969,Carbotte90,Carbotte_Marsiglio}.
   Within this formalism one assumes and then verifies that the pairing involves fermions in the vicinity of the FS, in which case the energy-dependent fermionic density of states can be approximated by its value at the chemical potential, and the dimensionless $s-$wave pairing  component reduces to a constant $\lambda <0$.  One can also verify that the corrections to the pairing vertex from renormalizations in the particle-hole channel are small in $\omega_L/\mu$  because in the
 processes that lead to the vertex renormalization, the fermions are forced to vibrate at phonon frequencies,
  far away from their own resonance frequencies (this is often termed the adiabatic limit~\cite{Scalapino1969,Carbotte90,Haslinger}).
 The ME formalism has been subsequently extended  to include the Coulomb repulsion.~\cite{Morel1962,Scalapino66,Scalapino1969,Carbotte90,McMillan1968,*Tolmachev}
    The bare Coulomb repulsion is stronger than the attraction due to phonon exchange, but it extends to frequencies of order $\mu$, while the electron-phonon interaction decays already at frequencies above $\omega_L$.  To first approximation, one then has to compare the electron-phonon attraction $\lambda$ with the effective Coulomb repulsion, renormalized by fermions with energies between $\mu$ and $\omega_L$.  These renormalizations are logarithmically singular in the particle-particle channel due to the presence of Coulomb logarithm, and they substantially reduce the strength of the Coulomb repulsion and reduce the dimensionless effective Coulomb interaction  to a constant $u^*$. It was argued~\cite{McMillan1968,*Tolmachev,Coleman_book} that $|\lambda| >u^*$, i.e., $s-$wave phonon-mediated superconductivity survives the Coulomb repulsion.       This last point has been verified in more sophisticated calculations~\cite{Morel1962,Gurevich1962,Scalapino66,Takada1980,Coleman_book,Ruhman2016,Ruhman2017},
 where the Coulomb and electron-phonon interactions were treated on equal footing, and the gap equation was solved in the full frequency range of order $\mu$ rather than in a narrow range of order $\omega_L$.  These calculations have found that the gap function $\Delta (\omega)$ avoids the strong Coulomb repulsion by changing sign as a function of frequency.
        This is qualitatively similar to how superconductivity emerges in quantum-critical electronic systems, where the pairing emerges from a nominally repulsive interaction mediated by a near-gapless collective boson, e.g, a spin fluctuation~\cite{Pines,Scalapino2012,Abanov2003,Mazin2008a,
        Kuroki,Chubukov2008,Chubukov_Maiti}. The gap function in such systems changes sign between the patches on a given FS, coupled most strongly by a critical boson, or between different Fermi pockets, like cuprate and Fe-based superconductors, respectively.  For systems with multiple bands, the sign change between the gaps on different FSs  may occur even if superconductivity is predominantly driven by the electron-phonon interaction.~\cite{Cohen1964,Grilli,Golubov1997,Trevisan2018}

At low electron density, when the characteristic phonon frequency becomes comparable, or even larger than $\mu$,   this consideration has to be modified by two reasons.  First, the pairing is no longer confined to the FS. Consequently, the equation for the pairing gap becomes a 2D integral equation in frequency and in momentum variation from $k_F$.  Simultaneously, the corrections to the pairing vertex grow and can no longer be rigorously neglected.~\cite{Grimaldi1995,Ikeda1992,Gorkov2016,Sadovskii2018}
Second, there is no longer a wide window for downturn renormalization of the Coulomb interaction, hence it is a priori unclear whether the attraction survives when both electron-phonon and Coulomb interaction are treated on equal footing~\cite{Sadovskii2018b}.
 Recent works proposed alternative sources of pairing in this regime, involving  plasmons~\cite{Takada1980,Ruhman2016,Ruhman2017}
or the polar coupling to phonons in ionic crystals.~\cite{Gurevich1962,Savary2017,Rowley2018,Lonzarich2018,Woelfle2018}

In this paper we revisit the electron-phonon pairing problem in a three-dimensional single-band system.
Our starting point is the modified Bardeen-Pines (BP)  model~\cite{BardeenPines}, which treats on equal footing the screened Coulomb repulsion and electron-phonon attraction.
The original BP model
 was introduced for a non-polar crystal, in which case the dressed phonon is an acoustic mode in the low-energy limit.
We consider semi-phenomenologically the modified  version of the BP model, in which we treat a boson as a gapped mode with a frequency $\omega_L$. The interaction potential in this modified BP model is the sum of a regularly screened Coulomb interaction and an interaction with a gapped dispersionless boson,
dressed by the Coulomb potential.
  The dressing endows the  frequency-dependent electron-boson interaction with momentum dependence.
 Below we will refer to the boson as the longitudinal  phonon, but in reality one should view our gapped mode  as an excitation  hybridized  between a longitudinal phonon and a plasmon~ (see e.g.,\cite{Ruhman2016,Ruhman2017,Mahan}).  In this respect, we note that our  semi-phenomenological model is similar to the one rigorously derived for strontium titanate in Ref.\cite{Ruhman2016} for a particular range of model parameters.    We do not restrict the
   parameters (the  ratio of $\omega_L/\mu$ in our case),  because our main goal is to understand how superconducting $T_c$ changes between the limits $\omega_L \ll \mu$ and $\omega_L \gg \mu$. We can justify our model in both limits (see Sec. II), but in between our model should be treated as a phenomenological model.

Given the complexity of the problem, here we consider only the attractive part of the interaction and address the issues of vertex corrections and the role of the states away from the FS.
   We analyze the interplay between the dressed electron-phonon interaction and the Coulomb repulsion in a separate paper.
  Our main goals here are to find the appropriate computation procedure in the anti-adiabatic limit, when $\omega_L$ is larger than $\mu$, and compute the onset temperature of the pairing  as a function of the ratio $\mu/\omega_L$. We label this temperature $T_c$ with the understanding that it coincides with the actual    superconducing $T_c$ only in the absence of phase fluctuations.~\cite{Randeria,SadeMelo1993,Chubukov2016}

     Within the ME formalism, the pairing is confined to the FS, and $T_c$ in the dilute limit vanishes because $\lambda$ is proportional to the density of states at the Fermi level, and the latter scales as $\sqrt{\mu}$. We show that the actual result is different because  at low carrier densities the pairing comes predominantly from electronic states far away from the FS. In essence,
 solving for the onset temperature of the pairing  at vanishing $\mu$ is analogous to solving for a bound state of two fermions in a vacuum. In a 3D  system a bound state forms when the attraction exceeds some threshold value. For a bare interaction between electrons and an optical phonon, $T_c$ would vanish at weak enough electron-phonon coupling. However, in the BP model, the electron-phonon interaction is dressed by a Coulomb potential, which becomes progressively less screened as $\mu$ decreases. Consequently, the effective electron-phonon coupling gets strongly enhanced and we show that it well exceeds the threshold value. As a result,  $T_c$ is non-zero even when $\mu =0$ and actually exceeds $\omega_L$ in this limit.
        We found analytically and confirmed numerically that in the dilute limit $T_c$ has a polynomial dependence on the phonon frequency $\omega_L$, $T_c\propto\omega_L\left(\frac{Ry}{\omega_L}\right)^\eta$, where $\text{Ry}$ is the Rydberg constant and $\eta \sim 0.2$.

 Phonon-mediated superconductivity in a system with small Fermi surface has been recently considered in Refs.~\cite{Aperis2018,Aperis2018b}.
  In their case the momentum-dependence of the interaction comes primarily from the momentum dependence of the bare electron-phonon coupling, which strongly favors $q=0$ scattering.
   The authors of~\cite{Aperis2018,Aperis2018b} also found that the states away from the Fermi level  contribute significantly to the pairing instability. The  distinction between our work and Refs.~\cite{Aperis2018,Aperis2018b} is that they computed $T_c$ numerically, while we obtain the analytical formula for $T_c$ at low fermion density. 

The paper is organized as follows. In the next section we introduce the BP model and obtain the equations for the fermionic self-energy and the s-wave component of the pairing vertex. In Sec.~\ref{section:results} we first solve for $T_c$  within the ME approximation and then investigate the validity of the ME formalism first in the regime  $\mu \sim \omega_L$ and then in the dilute limit $\mu \ll \omega_L$. We present the results of the full self-consistent analysis of $T_c$ as a function of $\mu/\omega_L$ and show that $T_c$ not only remains finite when $\mu =0$, but actually exceeds $\omega_L$.
We summarize our results in Sec.~\ref{section:summary}.

\begin{figure}
 \includegraphics[width=0.5\columnwidth]{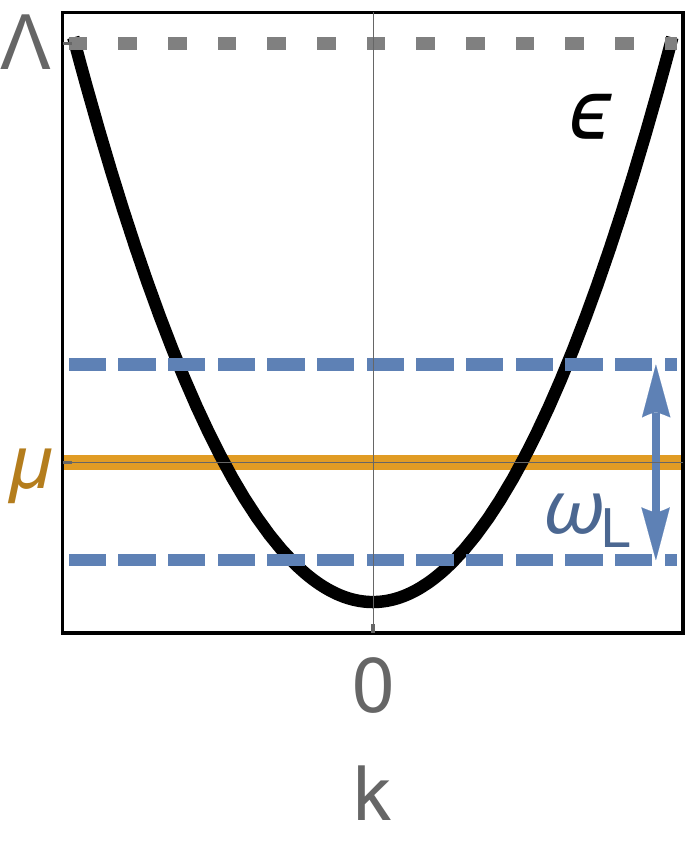}
\caption{{\it The energy scales.}  Black line --  the  parabolic electronic dispersion $\epsilon=k^2/2m$; orange line -- the chemical potential $\mu$; gray dashed line -- the bandwidth $\Lambda$; blue dashed lines -- the characteristic frequency of the optical phonon $\omega_L$.}
\label{fig:model}
\end{figure}

\section{The Model}

We consider an isotropic electron gas in 3 dimensions with dispersion $\xi(\mathbf k)=\mathbf{k}^2/2m-\mu$.
The electrons interact directly via the Coulomb repulsion and indirectly
via the exchange of phonons. In the most general case of a polar crystal, there are two optical phonon frequencies, the longitudinal one $\omega_{L}$
and the transverse one $\omega_{T}$. They are
related via the ionic plasma frequency $\omega_{p}$, as
$\omega_{L}^{2}=\omega_{T}^{2}+\omega_{p}^{2}$. The RPA expression
for the total interaction, which includes both the Coulomb and electron-phonon
contributions, is given in \cite{Ruhman2016,Mahan}
 and is a function of transferred momentum $q$ and transferred  frequency $\Omega$. Along the Matsubara axis $(\Omega_n = 2\pi nT$) it is expressed as
\begin{equation}
V_{n}\left(\mathbf{q}\right)=\frac{W_{0}\left(\mathbf{q}\right)}{\varepsilon_{\infty}\frac{\Omega_{n}^{2}+\omega_{L}^{2}}{\Omega_{n}^{2}+\omega_{T}^{2}}-\Pi\left(\mathbf{q},\Omega_{n}\right)W_{0}\left(\mathbf{q}\right)}\label{V_general}
\end{equation}
Here
 $W_{0}\left(\mathbf{q}\right)=4\pi e^{2}/q^{2}$ is the
bare Coulomb repulsion, $\Pi\left(\mathbf{q},\Omega_{n}\right)$ is the Lindhard
function, and
 $\varepsilon_{\infty}$ is the value of the dielectric constant in the large-frequency limit.
  The  $\varepsilon_{\infty}$  is related to the zero-frequency value
$\varepsilon_{0}$ by $\varepsilon_{\infty} = \varepsilon_{0} - \omega_{p}^{2}/\omega_{T}^{2}$
 (or, equivalently,
$\omega_{p}^{2}=\left(\varepsilon_{0}-\varepsilon_{\infty}\right)\omega_{T}^{2}$).

This expression comprises two well-understood regimes. In the case
of a an insulating ionic crystal, $\Pi$ can be set to zero, and the
effective electron-electron interaction is the Coulomb interaction
 screened  by
 the polar optical phonons:
\begin{equation}
V_{n}^{\mathrm{polar}}\left(\mathbf{q}\right)=\frac{4\pi e^{2}}{\varepsilon_{\infty}q^{2}}\left[1-\frac{\omega_{L}^{2}-\omega_{T}^{2}}{\Omega_{n}^{2}+\omega_{L}^{2}}\right]\label{V_polar}
\end{equation}

Conversely, in a non-polar crystal, we can set $\omega_{T}=0$ and
$\varepsilon_{\infty}=1$ to obtain the BP expression
for the dynamical interaction,
 which vanishes in the static limit. At $\Omega_n \ll v_F q$, $\Pi\left(\mathbf{q},\Omega_{n}\right)$ can be approximated
  by its static form, and the BP potential becomes
\begin{equation}
\label{eq:VBP}
V^{BP}_n(\mathbf q)=\frac{4\pi e^2}{q^2 \varepsilon (q)}
\left[1-\frac{\omega_{q}^2}{\Omega_n^2+\omega_{q}^2}\right],
\end{equation}
We set $q \ll k_F$ and  used that at small $q$
,
 $\Pi W_{0}\approx1-\varepsilon\left(q\right)$,
where $\varepsilon(q)=1+\frac{\kappa^2}{q^2}$ and $\kappa^{2}=4\pi e^2 N(\mu)$ is the
Thomas-Fermi screening momentum
  expressed via the  density of states $N(\mu)$.
  The frequency $\omega_q$ in this approximation is
the longitudinal phonon frequency $\omega_L$, renormalized by the electronic polarization:
  $\omega_{q}^2=\omega_{L}^2/\varepsilon\left(q\right) = \omega^2_L q^2/(q^2 + \kappa^2)$.  It becomes an acoustic phonon in the long wave-length
limit. Note that $\omega_L$ in the BP model is  the same as the ionic plasma frequency $\omega_{p}$, because $\omega_T=0$.

 The authors of \cite{Ruhman2016} considered a polar crystal with a finite density of conduction electrons and
 obtained the effective $V_n (q)$ by integrating out the phonon degrees of freedom. For parameters relevant to  SrTiO$_{3}$
  they obtained the effective interaction similar to Eq. \ref{eq:VBP}, but with $\epsilon_0$ instead of $\epsilon (q)$ and with
  a $q-$independent  renormalized electronic plasma frequency instead of $\omega_q$.
  Because our interest
is more general than polar SrTiO$_{3}$,
we focus instead on the BP
limit, Eq.~\eqref{eq:VBP}.
This will allow us to interpolate between the limits $\mu\gg\omega_L$,
as appropriate for standard superconductors and $\mu\ll\omega_L$, as appropriate for dilute superconductors.

Importantly, in the reminder of the paper, we will
neglect the renormalization of the optical mode $\omega_{L}$ and
continue to use the bare frequency $\omega_{L}$ instead of $\omega_{q}$. The reasoning for this approximation is the following. In the limit of $\mu\gg\omega_L$, the relevant electronic processes are those around the Fermi level, and the renormalization of $\omega_L$ by $\varepsilon(q)$ can be neglected within the leading logarithmic approximation (see Appendix~\ref{appendix:interaction}).
In the opposite limit of very dilute systems $\mu\ll\omega_L$, the screening is negligible and $\varepsilon (q) \approx 1$.
In between the two limits,  the difference between
$\omega_L$ and $\omega_q$ is not negligible, but does not affect the trend of $T_c$ variation with $\mu/\omega_L$.

The effective electron-electron interaction $V^{BP}_n(\mathbf q)$  consists of an instantaneous repulsive screened Coulomb term and a retarded attractive electron-phonon term. Given the complexity of this interaction, in  this work we focus on the attractive part of the effective interaction:

\begin{equation}
\label{eq:Veff}
V_n^{\text{e-ph}}(q)=-\frac{4\pi e^2}{q^2+\kappa^2}\left[\frac{\omega_L^2}{\Omega_n^2+\omega_L^2}\right].
\end{equation}

We will use the Nambu-Gor'kov formalism in which the pairing vertex is the $\hat \tau_1$ piece of the matrix electronic self-energy $\hat\Sigma_n(\mathbf k)$ ($\hat \tau_i$ are Pauli matrices). The matrix Dyson equation,  relating the Green's function $\mathcal{\hat G}^{-1}_n(\mathbf{k})$ to $\hat\Sigma_n(\mathbf k)$, is given by
\begin{equation}
\label{eq:fullG}
\mathcal{\hat G}^{-1}_n(\mathbf{k})=i\nu_n \hat{\tau}_0 - \xi(\mathbf k)\hat{\tau}_3-\hat\Sigma_n(\mathbf{k})
\end{equation}
where  $\omega_n=(2n+1)\pi T$ are fermionic Matsubara frequencies. We compute the fermionic self-energy $\hat\Sigma_n(\mathbf k)$ self-consistently by  expressing it as a convolution of the interaction
$V^{\text{e-ph}}_{n-n'}(|\mathbf{k}-\mathbf{k}'|)$ and the full fermionic propagator $\mathcal{\hat G}_{n'}(\mathbf{k}')$:
\begin{align}
\label{eq:selfenergy}
\hat\Sigma_n(\mathbf{k})&=-T\sum_{\mathbf{k}',n'}V^{\text{e-ph}}_{n-n'}(|\mathbf{k}-\mathbf{k}'|)\hat{\tau}_3 \mathcal{\hat G}_{n'}(\mathbf{k}')\hat{\tau}_3\\\nonumber
&=i\omega_n\left[1- Z_n(\mathbf{k})\right]\hat{\tau}_0 +\chi_n(\mathbf k)\hat{\tau}_3+\phi_n(\mathbf k)\hat{\tau}_1
\end{align}
In \eqref{eq:selfenergy} the self-energy
is decomposed into the two 'normal' components, the mass renormalization function $Z_n(\mathbf k)$ and  energy shift $\chi_n(\mathbf k)$, and the 'anomalous' component (the pairing vertex) $\phi_n(\mathbf k)$ (Ref.~\onlinecite{Scalapino66}).
Imposing self-consistency in Eqs.~\eqref{eq:fullG}-\eqref{eq:selfenergy} we obtain  a set of coupled equations for the functions $Z_n(\mathbf k)$, $\chi_n(\mathbf k)$ and  $\phi_n(\mathbf k)$, \begin{widetext}
\begin{align}
\label{eq:Zfull_0}
Z_n({\bf k})-1&=
-T\frac{1}{\omega_n}\sum_{n'}
\int \frac{d^3 {\bf k'}}{(2\pi)^3}
V^{\text{e-ph}}_{n-n'}(|\mathbf{k}-\mathbf{k}'|)
\frac{\omega_{n'} Z_{n'}({\bf k}')}{\left[\omega_{n'}Z_{n'}(  {\bf k}')\right]^2+\left[\xi(\mathbf k')+\chi_{n'}({\bf k}')\right]^2}\\
\chi_n({\bf k})&=
T\sum_{n'}
\int \frac{d^3 {\bf k'}}{(2\pi)^3}
V^{\text{e-ph}}_{n-n'}(|\mathbf{k}-\mathbf{k}'|)
\frac{\xi(\mathbf k')+\chi_{n'}({\bf k}')}{\left[\omega_{n'}Z_{n'}({\bf k}')\right]^2+\left[\xi(\mathbf k')+\chi_{n'}({\bf k}')\right]^2}\\
\label{eq:phifull_0}
\phi_n({\bf k})&=
-T\sum_{n'}\int \frac{d^3 {\bf k'}}{(2\pi)^3} V^{\text{e-ph}}_{n-n'}(|\mathbf{k}-\mathbf{k}'|)
\frac{\phi_{n'}({\bf k}')}{\left[\omega_{n'}Z_{n'}({\bf k}')\right]^2+\left[\xi(\mathbf k')+\chi_{n'}({\bf k}'
)\right]^2}.
\end{align}
\end{widetext}
We linearized the equations with respect to the pairing vertex $\phi_n(\mathbf{k})$ to compute the superconducting temperature. The computations are simplified by the fact that the interaction  $V^{\text{e-ph}}_{n-n'}(|\mathbf{k}-\mathbf{k}'|)$ is factorized between momentum and frequency dependencies: $V^{\text{e-ph}}_{n-n'}(|\mathbf{k}-\mathbf{k}'|)= u_{n-n'}\mathcal{V}(|\mathbf{k}-\mathbf{k}'|)$, where
\begin{align}
\label{eq:unn'}
u_{n-n'}&=\frac{\omega^2_L}{\omega^2_L +|\omega_n-\omega_{n'}|^2};\\\nonumber
\mathcal{V}(|\mathbf{k}-\mathbf{k}'|) &= - \frac{4 \pi e^2_0}{|{\bf k} - {\bf k}'|^2 + \kappa^2}.
\end{align}
For a rotationally-isotropic fermionic dispersion, which we consider here, $\mathcal{V}(|\mathbf{k}-\mathbf{k}'|)$ and the pairing vertex $\phi_n({\bf k})$ can be expanded in partial components for different angular momentum $l$. The fermionic $Z_n ({\bf k})$ and the energy shift $\chi_n ({\bf k}) $ depend on the magnitude of momentum $|{\bf k}|$, but not on its direction, and are expressed in terms of the s-wave component of the interaction $v(k,k')=\frac{1}{2\pi}\int_{-1}^1 dx \mathcal{V}(\sqrt{k^2+k^{'2}-2kk'x})$. The equations for different partial harmonics of $\phi_n ({\bf k})$  decouple, and we focus on the  $s-$wave solution for the pairing vertex $\phi^s_n ({\bf k}) \equiv \phi_n (k)$.
 The momentum component of the pairing interaction  for $ \phi_n (k)$ is the same for $Z_n (k)$ and $\chi_n (k)$.

Below we rescale all variables by the characteristic phonon frequency, i.e., introduce the rescaled  temperature $\bar T=\frac{T}{\omega_L}$, rescaled Matsubara frequency $\bar\omega_n=\frac{\omega_n}{\omega_L}$, rescaled energy $\bar\epsilon=\frac{k^2/2m}{\omega_L}$, rescaled $\bar\kappa^2=\frac{\kappa^2/2m}{\omega_L}$, rescaled chemical potential $\bar\mu=\frac{\mu}{\omega_L}$, rescaled bandwidth $\bar\Lambda=\frac{\Lambda}{\omega_L}$ (the cutoff in momentum space), and $\bar\rho=\frac{\text{Ry}}{\omega_L}$, the rescaled variable of the Rydberg energy $\text{Ry}=\frac{m e^4}{2}=13.6$ eV. In these notations, the set of coupled equations for $Z_n (\bar \epsilon)$, ${\bar \chi}_n ({\bar \epsilon})$, and $s-$wave pairing potential $\phi_n ({\bar \epsilon})$ are
\begin{widetext}
 \begin{align}
 \label{eq:Zfull}
Z_n(\bar\epsilon)-1&=\bar T\frac{1}{\bar\omega_n}\sum_{n'}u_{n-n'}
\int_0^{\bar\Lambda} d\bar\epsilon'N(\bar\epsilon')v(\bar\epsilon,\bar\epsilon')\frac{\bar\omega_{n'} Z_{n'}(\bar\epsilon')}{\left[\bar\omega_{n'}Z_{n'}(\bar\epsilon')\right]^2+\left[\bar\epsilon'-\bar\mu+\bar\chi_{n'}(\bar\epsilon')\right]^2}\\
\label{eq:chifull}
\bar\chi_n(\bar\epsilon)&=-\bar T\sum_{n'}u_{n-n'}\int_0^{\bar\Lambda} d\bar\epsilon'N(\bar\epsilon')v(\bar\epsilon,\bar\epsilon')\frac{\bar\epsilon'-\bar\mu+\bar\chi_{n'}(\bar\epsilon')}{\left[\bar\omega_{n'}Z_{n'}(\bar\epsilon')\right]^2+\left[\bar\epsilon'-\bar\mu+\bar\chi_{n'}(\bar\epsilon')\right]^2}\\
\label{eq:phifull}
\phi_n(\bar\epsilon)&=\bar T\sum_{n'}u_{n-n'}\int_0^{\bar\Lambda} d\bar\epsilon'N(\bar\epsilon')v(\bar\epsilon,\bar\epsilon')\frac{\phi_{n'}(\bar\epsilon')}{\left[\bar\omega_{n'}Z_{n'}(\bar\epsilon')\right]^2+\left[\bar\epsilon'-\bar\mu+\bar\chi_{n'}(\bar\epsilon')\right]^2},
\end{align}
\end{widetext}
where $N({\bar \epsilon})=\sqrt{\bar\epsilon}$ is the density of states and the s-wave component of the interaction is
\begin{equation}
\label{eq:vpp}
v(\bar\epsilon,\bar\epsilon')=\frac{\sqrt{{\bar\rho}}}{2\pi\sqrt{\bar\epsilon \bar\epsilon'}}\log \left[\frac{(\sqrt{\bar\epsilon}+\sqrt{\bar\epsilon'})^2+\bar\kappa^2}{(\sqrt{\bar\epsilon}-\sqrt{\bar\epsilon'})^2+\bar\kappa^2}\right].
\end{equation}
For further details of the derivation of these equations see Appendix~\ref{appendix:eqsderivation}.
We emphasize that the chemical potential $\bar\mu$ in Eqs.~\eqref{eq:Zfull}-\eqref{eq:phifull} is defined at $\bar T_c$ rather than at $T=0$, where it is equal to the (rescaled) Fermi energy. The value of ${\bar \mu}$ at $\bar T_c$ is smaller than at $T=0$.
We will obtain $\bar T_c$ as a function of ${\bar \mu}$ at $\bar T_c$ and set minimal ${\bar \mu} =0$.  The Fermi energy for such minimal ${\bar \mu}$ is still  finite.
Throughout this work we will set $\omega_L=100$ meV and $\bar\Lambda=100$, unless otherwise stated.
The numerical solution of the integral equations were computed by using the Nystr\"{o}m method with a N-point Gauss-Legendre rule ($100<N<200$).

In the following section we revisit the computation of the electronic self-energy within the Migdal-Eliashberg (ME) approximation.
In this approximation, the temperature variation of ${\bar \mu}$ is neglected, and the integration over ${\bar \epsilon}'$ is confined to the FS, in which case $N({\bar \epsilon}')v(\bar \epsilon,\bar\epsilon')$ can be approximated by its value at the Fermi energy. Then the electronic self-energy is independent of ${\bar \epsilon}$, i.e. $\hat\Sigma_n (\bar \epsilon)=\hat\Sigma_n$, and the energy shift ${\bar \chi}_n ({\bar \epsilon})$ reduces to a constant and can be absorbed into the chemical potential. This eliminates the variable $\bar\epsilon$ from the Eliashberg equations and reduces them to integral equations only in Matsubara frequency. The ME approximation is fully justified when the characteristic frequency $\omega_L$ is small compared to the Fermi energy of the electrons,  i.e., ${\bar \mu} \gg 1$, as corrections are small in $1/{\bar \mu}$. [Corrections to the fermion-boson vertex, which we did not include into Eqs. (\ref{eq:Zfull_0})-(\ref{eq:phifull_0}), are also small in  $1/{\bar \mu}$.] We investigate the validity of  ME approximation as ${\bar \mu}$ gets smaller. We show that the key new effect at small ${\bar \mu}$ is that one cannot restrict the integration over ${\bar \epsilon}'$ to the FS and approximate $N({\bar \epsilon})v(\bar \epsilon,\bar\epsilon')$ by its value at the Fermi energy. Instead one should solve Eqs. (\ref{eq:Zfull_0})-(\ref{eq:phifull_0}) as integral equations in both Matsubara frequency $\omega_n$ and the energy variable $\epsilon$. Other deviations from ME approximation are at most $O(1)$ and are not fundamentally relevant.

\begin{figure*}
 \includegraphics[width=\textwidth]{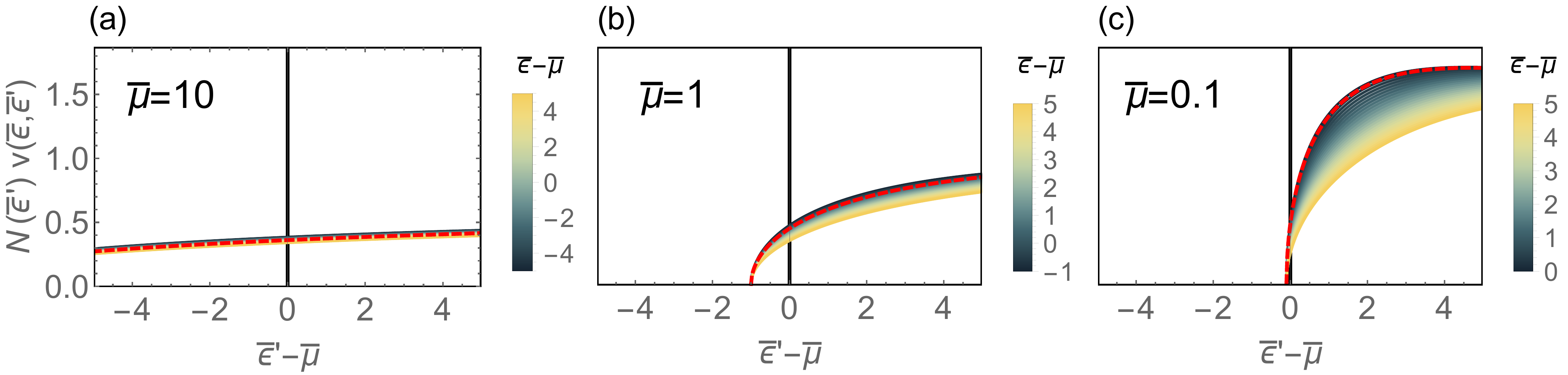}
\caption{{\it The effective dimensionless s-wave coupling constant}.
The dimensionless coupling constant is the  product of the  density of states $N(\bar\epsilon')=\sqrt{\bar\epsilon'}$ and the  s-wave interaction $v(\bar\epsilon,\bar\epsilon')$, Eq.~\eqref{eq:vpp}.  The product $N(\bar\epsilon') v(\bar\epsilon,\bar\epsilon')$ is plotted
 for several  ${\bar \epsilon}$
in a $\pm 5\omega_L$ energy window of ${\bar \epsilon}'$ around the rescaled chemical potential $\bar\mu=\frac{\mu}{\omega_L}$.
  Panels (a)-(c) are for $\bar\mu=10$,  $\bar\mu=1$, and $\bar\mu=0.1$, respectively. The dimensionless coupling
   acquires a strong energy dependence on $(\bar\epsilon,\bar\epsilon')$ as $\bar\mu$ is reduced.
    The dashed red line corresponds to $\bar\epsilon=\bar\mu$. The interception of this lines with $\bar\epsilon'-\bar\mu=0$ sets the ME coupling constant $\lambda$ [Eq.\eqref{eq:lambda}]. }
\label{fig:integrand}
\end{figure*}

\section{The results}
\label{section:results}
\subsection{Standard ME approximations}

To be more precise, the ME theory  is based on two approximations, both  justified by the smallness of the characteristic frequency $\omega_L$ compared to the Fermi energy of the electrons:
\begin{enumerate}
\item The energy dependent $N(\bar\epsilon')v(\bar\epsilon,\bar\epsilon')$ is replaced by a constant
$N(\bar\mu)v(\bar\mu,\bar\mu)$ as relevant scattering is assumed to be restricted to frequencies of order $\omega_L$. In Fig.~\ref{fig:integrand}(a) we plot $N(\bar\epsilon')v(\bar\epsilon,\bar\epsilon')$  in a $\pm 5 \omega_L$ energy window around the chemical potential for $\bar\mu=10$. We see that $N(\bar\epsilon')v(\bar\epsilon,\bar\epsilon')$ is indeed close to $N(\bar\mu)v(\bar\mu,\bar\mu)$.
\item
The integration over $\bar\epsilon' - \bar\mu$ in the r.h.s. of Eqs. (\ref{eq:Zfull}), (\ref{eq:chifull}), and  (\ref{eq:phifull}) is extended to infinite limits, from $-\infty$ to $+\infty$. Within this approximation,   ${\bar \chi}_{n}=0$, and the r.h.s. of the equation for the quasiparticle residue $Z_n$ becomes independent of $Z_{n'}$, i.e., it could be computed using  free-fermion propagators.
\end{enumerate}

\begin{figure}
 \includegraphics[width=\columnwidth]{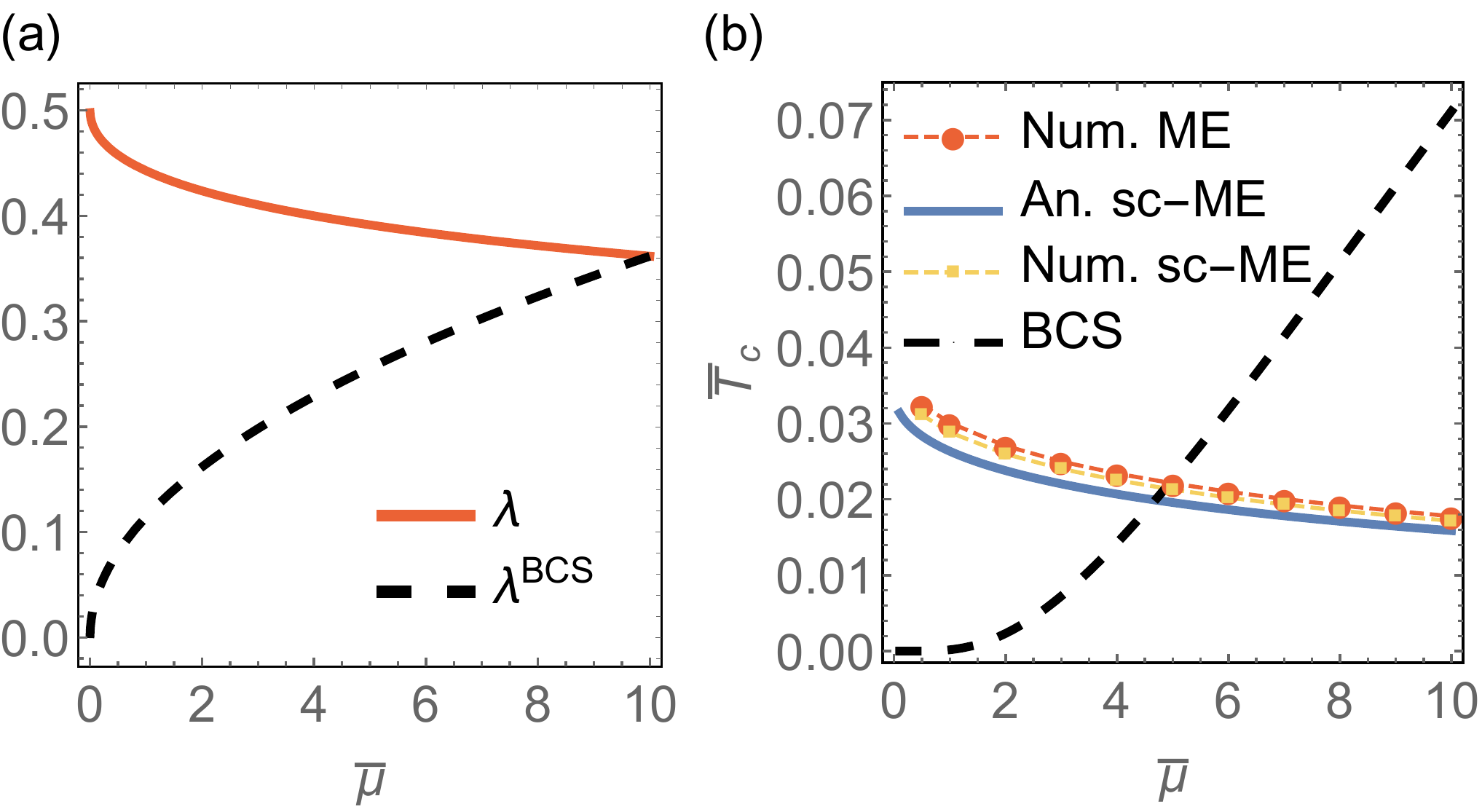}
\caption{\emph{The pairing within the ME formalism}.
(a) The dimensionless coupling constant.  Red line -- the dependence  of $\lambda$ from Eq.~\eqref{eq:lambda} on
 $\bar\mu$; black dashed line -- the BCS coupling $\lambda^{\text{BCS}}=\sqrt{\bar\mu}v(\bar\mu=10,\bar\mu=10)$.
(b) Critical temperature $\bar T_c$ vs $\bar\mu$.
  Red dots --  the numerical solution of the ME equation \eqref{eq:gapeq_delta};
 yellow squares -- the numerical solution of the ME equation ~\eqref{eq:gapeq_fs_*} in the small coupling limit; blue line --
   the analytic solution of the ME equation in the weak coupling limit (Eq.~\eqref{eq:tc_elias});
    black dashed line -- the  canonical  BCS expression for $T_c$,  Eq.~\eqref{eq:tc_bcs}.
   }
\label{fig:pairing_fs}
\end{figure}

Within these two approximations the set of Eliashberg equations for $\bar T_c$ reduces to two integral equations for the mass renormalization $Z_n$ and the pairing vertex $\phi_n$:
\begin{align}
\label{eq:gapeq_fs2}
Z_n&= 1 + \lambda \frac{\pi \bar T}{\bar\omega_n}\sum_{n'}\frac{\sgn (\bar\omega_{n'})}{1+|\bar\omega_{n'}-\bar\omega_n|^2}\\
\label{eq:gapeq_fs}
\phi_n &=\lambda \pi \bar T\sum_{n'}\frac{1}{1+|\bar\omega_{n'}-\bar\omega_n|^2}\frac{\phi_{n'}}{|\bar\omega_{n'}| Z_{n'}}.
\end{align}
Here
\begin{equation}
\label{eq:lambda}
\lambda\equiv N(\bar\mu)v(\bar\mu,\bar\mu)=\frac{1}{2\pi}\sqrt{\frac{\text{Ry}}{\mu}}\log \left[1+\pi\sqrt{\frac{\mu}{\text{Ry}}}\right],
\end{equation}
is a dimensionless coupling constant. It increases monotonically when the chemical potential is reduced and in the limit ${\bar \mu} \to 0$ reaches $\lambda({\bar \mu}\rightarrow 0)=0.5$.
This increase is a consequence of the reduced Thomas-Fermi screening length of the electron gas at smaller electronic densities. We show $\lambda$  as a function of ${\bar \mu}$ in Fig.~\ref{fig:pairing_fs}(a) (red line).

The set of the two  equations (\ref{eq:gapeq_fs2}) and (\ref{eq:gapeq_fs}) can be reduced to one integral equation for $T_c$  by introducing the superconducting order parameter $\Delta_n\equiv\frac{\phi_n}{Z_n}$, expressing $\phi_n$ via $\Delta_n$, and substituting the explicit form of $Z_n$ from  Eq. (\ref{eq:gapeq_fs2}). This yields a 1D integral equation for $\Delta_n$:
\begin{equation}
\label{eq:gapeq_delta}
\Delta_n  =\lambda \pi \bar T\sum_{n'}\frac{1}{1+|\bar\omega_{n'}-\bar\omega_n|^2}\left(\frac{\Delta_{n'}}{|\bar\omega_{n'}|} - \frac{\Delta_{n}}{|\bar\omega_{n'}|} \sgn \omega_{n'}\right).
\end{equation}
This equation has been extensively studied in the literature.
The $\bar T_c^{\text{ME}}$, obtained by numerically solving Eq. (\ref{eq:gapeq_delta}),  is shown in Fig.~\ref{fig:pairing_fs}(b) (red circles)  as a function of the chemical potential $\bar\mu$, which enters the gap equation through $\lambda$. As the chemical potential is reduced, the critical temperature steadily increases. This growth is  expected because the coupling constant $\lambda$ increases with decreasing $\mu$  (see Fig. ~\ref{fig:pairing_fs}(a)).

At small $\lambda$, the quasiparticle residue $Z_n$ does not depend on $n$ to first order in $\lambda$: $Z_n = 1 + \lambda + \mathcal{O}(\lambda^2)$.
If we use this $Z_n$, we simplify the  equation for $T_c$ even further, to
\begin{equation}
\label{eq:gapeq_fs_*}
\phi_n=\lambda^* \pi \bar T\sum_{n'}\frac{1}{1+|\bar\omega_{n'}-\bar\omega_n|^2}\frac{1}{|\bar\omega_{n'}|}\phi_{n'}
\end{equation}
where $\lambda^*=\frac{\lambda}{1+\lambda}$.
This equation can be solved analytically, again by expanding in the coupling~\cite{Karakozov75,Dolgov05,Wang13,Marsiglio18}
and yields
\begin{equation}
\label{eq:tc_elias}
 \bar T_c^{\text{An.sc-ME}}=\frac{1.13}{\sqrt{e}} e^{-\frac{1}{\lambda^*}}=0.7 e^{-\frac{1+\lambda}{\lambda}}
\end{equation}
We plot this $\bar T_c^{\text{An.sc-ME}}$ as a function of ${\bar \mu}$ as a blue line in  Fig.~\ref{fig:pairing_fs}(b). The agreement between the exact $\bar T_c^{\text{ME}}$ and the analytical $ \bar T_c^{\text{An.sc-ME}}$ is quite good for $\lambda$ roughly below $0.5$.
The agreement becomes even better if we extract $\bar T_c$ from  Eq. (\ref{eq:gapeq_fs_*}) by solving it numerically instead of using the approximate analytical result. We show this $\bar T_c^{\text{sc-ME}}$ by yellow squares in Fig. \ref{fig:pairing_fs}(b).

For comparison with the full solution of Eqs. (\ref{eq:Zfull}), (\ref{eq:chifull}), and  (\ref{eq:phifull}) later in the paper, it is instructive
 to modify the second approximation  used to obtain the Eliashberg equations (\ref{eq:gapeq_fs2})
  and (\ref{eq:gapeq_fs}) and keep the integration over $\bar\xi'=\bar\epsilon' - \bar\mu$ in finite limits, but use the free-fermion Green's functions instead of dressed fermions to compute $\bar Z_n$ and ${\bar \chi}_n$.   Within this approximation the set (\ref{eq:Zfull}), (\ref{eq:chifull}), and  (\ref{eq:phifull}) reduces to
\begin{align}
\label{eq:Z0}
Z_n^{(0)}&=1+\lambda \bar T\frac{1}{\bar\omega_n}\sum_{n'}u_{n-n'}\bar\omega_{n'}
\int_{-\bar\mu}^{\bar\Lambda-\bar\mu} d\bar\xi'\frac{1}{\bar\omega_{n'}^2+\bar\xi^{'2}}\\
\label{eq:chi0}
\bar\chi_n^{(0)}&=-\lambda \bar T\sum_{n'}u_{n-n'}\int_{-\bar\mu}^{\bar\Lambda-\bar\mu} d\bar\xi'\frac{\bar\xi'}{\bar\omega_{n'}^2+\bar\xi^{'2}}\\
\label{eq:phi0}
\phi_n&= \lambda \bar T\sum_{n'}u_{n-n'}\int_{-\bar\mu}^{\bar\Lambda-\bar\mu} d\bar\xi' \frac{\phi_{n'}}{\left[\bar\omega_{n'}Z_{n'}^{(0)}\right]^2+
\left[\bar\xi'+ \chi^{(0)}_{n'}\right]^2}
\end{align}
We introduce the superscript $(0)$ to specify that $Z_n$ and $\bar\chi_n$  are obtained with free-fermion propagators. The integration over  $\bar\xi'$ can again  be performed analytically, and the set (\ref{eq:Z0})-(\ref{eq:phi0}) reduces to
\begin{align}
\label{eq:Z0fs}
Z_n^{(0)}&=1+\lambda \frac{\bar T}{\bar\omega_n}\sum_{n'}u_{n-n'}\zeta_{n'}^{(0)}\\
\label{eq:chi0fs}
\bar\chi_n^{(0)}&=-\lambda \bar T \sum_{n'}u_{n-n'}\eta_{n'}^{(0)}\\
\label{eq:phi0fs}
\phi_n&=\lambda \bar T\sum_{n'}u_{n-n'}\phi'_n\zeta_{n'}
\end{align}
where the functions $\zeta_{n'}^{(0)}$, $\eta_{n'}^{(0)}$ and $\zeta_{n'}$ are given by
\begin{align}
\zeta_{n'}^{(0)}&=\frac{1}{\bar\omega_{n'}}\left[\arctan\left(\frac{\Lambda-\mu}{\omega_{n'}}\right)+\arctan\left(\frac{\mu}{\omega_{n'}}\right)\right]\\
\eta_{n'}^{(0)}&=\frac{1}{2}\log\frac{\omega_{n'}^{2}+(\Lambda-\mu)^2}{\omega_{n'}^{2}+\mu^2}\\
\zeta_{n'}&=\frac{1}{\bar\omega_{n'} Z_{n'}}\left[\arctan\left(\frac{\Lambda-\mu+\chi^{(0)}_{n'}}{\omega_{n'}Z_{n'}^{(0)}}\right)\right. \\\nonumber
&\qquad\qquad\quad \left. +\arctan\left(\frac{\mu-\chi^{(0)}_{n'}}{\omega_{n'} Z_{n'}^{(0)}}\right)\right].
\end{align}
In the infinite bandwidth limit ($\Lambda\gg\mu\rightarrow\infty$), $\zeta_{n'}^{(0)}=\frac{\pi}{|\bar\omega_{n'}|}$, $\eta_{n'}^{(0)}=0$, and $\zeta_{n'}=\frac{\pi}{|\bar\omega_{n'} Z_{n'}|}$. Then we recover the conventional Eliashberg equations \eqref{eq:gapeq_fs2}-\eqref{eq:gapeq_fs}.

We wrap up this Section with a few words on the critical temperature within the "conventional" BCS formalism. In the conventional scenario, an effective electron-phonon interaction $v_{\text{eff}}$ is taken to be independent of the chemical potential $\mu$. The density of states of the electron gas in 3 dimensions scales as $\sqrt{\bar\mu}$, hence the dimensionless coupling constant $\lambda^{\text{BCS}}=\sqrt{\bar\mu}v_{\text{eff}}$.  We plot $\lambda^{\text{BCS}}$ as a dashed black line in figure~\ref{fig:pairing_fs}(a), matching it with our $\lambda$ at $\bar\epsilon= \bar\epsilon'=10$. As $\bar\mu$ decreases, $\lambda^{\text{BCS}}$ decreases, hence the critical temperature
\begin{equation}
\label{eq:tc_bcs}
\bar T_c^{\text{BCS}}=1.13  e^{-\frac{1}{\sqrt{\bar\mu}v_{\text{eff}}}}
\end{equation}
gets strongly reduced. We plot this $\bar T_c^{\text{BCS}}$ as a black dashed line  in Fig. ~\ref{fig:NS_fsrestriction}(b). We see that it is quite different from the actual $\bar T_c^{\text{ME}}$, which tends to a constant as $\bar\mu\rightarrow 0$.

\subsection{Validity of the ME approximations}
\label{subsection:validity}
We turn now to the analysis of the validity of the two
ME approximations as the system  moves away from the  $\bar\mu\gg 1$ regime  into the regime where the chemical potential is comparable or even smaller than the characteristic phonon frequency, $\bar\mu\ll 1$.

\begin{figure}
 \includegraphics[width=\columnwidth]{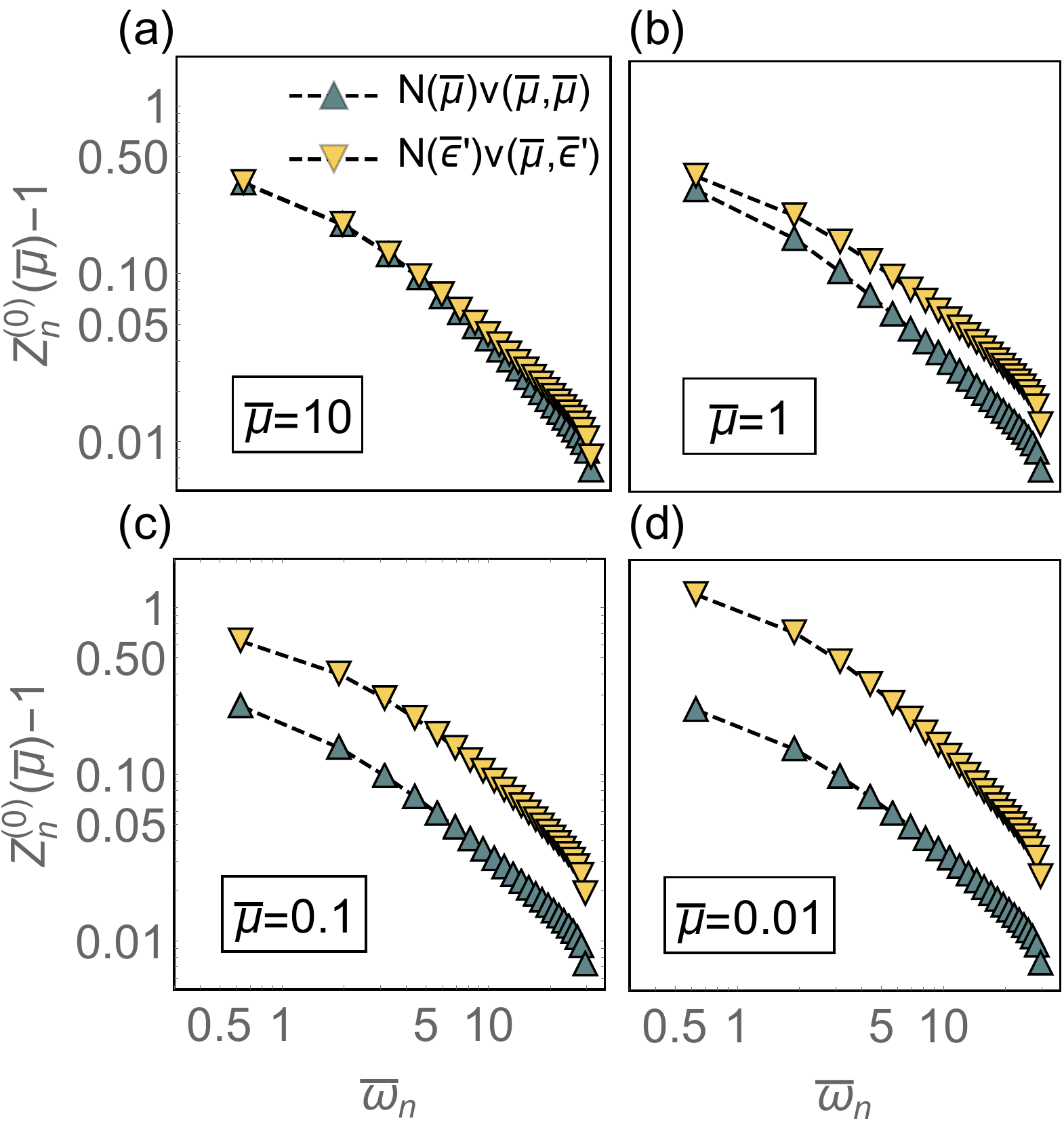}
\caption{\emph{The analysis of the restriction to the FS for $Z_{n}^{(0)}(\bar\mu)$.}
Panels (a)-(d) - the plots of $Z_{n}^{(0)}(\bar\mu)-1$ vs $\bar\omega_n$ in a log-log scale, computed with FS restriction (blue up-triangles) and without it (yellow down-triangles) for different ${\bar \mu}$.  (a) $\bar\mu=10$ ($\bar\kappa^2\sim 50$), (b) $\bar\mu=1$ ($\bar\kappa^2\sim 15$), (c) $\bar\mu=0.1$ ($\bar\kappa^2\sim 5$) and (c) $\bar\mu=0.01$ ($\bar\kappa^2\sim 1.5$). In all panels $\bar T=0.2$.}
\label{fig:NS_fsrestriction}
\end{figure}

The evolution of the energy dependent coupling constant $N(\bar\epsilon')v(\bar\epsilon,\bar\epsilon')$, which appears in the r.h.s. of Eqs.~\eqref{eq:Zfull}-\eqref{eq:phifull}, is shown in figure~\ref{fig:integrand} for various ${\bar \mu}$.
We put $\bar\epsilon'-\bar\mu$ along the horizontal axis and set it to be in the window $|\bar\epsilon'-\bar\mu|<5$. For each value of ${\bar \epsilon}' - {\bar \mu}$ we analyze the variation of $N(\bar\epsilon')v(\bar\epsilon,\bar\epsilon')$ with $\bar \epsilon -\bar \mu$ by analyzing how much $N(\bar\epsilon')v(\bar\epsilon,\bar\epsilon')$ differs from $N(\bar\epsilon')v(\bar\mu,\bar\epsilon')$, which we present as a dashed red line.  The value of $N(\bar\epsilon')v(\bar\mu,\bar\epsilon')$ at $\bar\epsilon'=\bar\mu$  is the coupling constant $\lambda$ used in the ME analysis. We see from the figure that $N(\bar\epsilon')v(\bar\epsilon,\bar\epsilon')$ is close to $\lambda$ at large $\bar\mu$, but develops a strong energy dependence when ${\bar \mu}$ becomes small. Consequently, at small $\bar\mu$,  the energy integration in Eqs.~\eqref{eq:Zfull}-\eqref{eq:phifull} with and without the restriction to the vicinity of the Fermi level would result in very different normal and anomalous self-energies. We illustrate this in Fig.~\ref{fig:NS_fsrestriction}, where we compare the results  for the quasiparticle residue $Z_n (\bar \mu)$, computed in two ways. For definiteness we compare the values of $Z^{(0)}_n (\bar \mu)$,  which, we remind, we compute with free fermions, but  integrating over $\bar\xi'$ in  finite limits.  We see from Fig. \ref{fig:NS_fsrestriction}(a) that at large ${\bar \mu}$, $Z_n^{(0)}(\bar\mu)-1$ obtained with an without restriction to the FS are essentially identical. However, for smaller ${\bar \mu}$, $Z^{(0)}(\bar\mu)-1$, obtained by keeping the full energy dependence in the coupling constant,  substantially increases, while  $Z^{(0)}(\bar\mu)-1$ obtained in FS-restricted calculations remains essentially unchanged.  We show this in panels (b)-(d) in  Fig. \ref{fig:NS_fsrestriction}. The ratio of the two $Z_n^{(0)}(\bar\mu)-1$ reaches 5 for $\bar\mu=0.01$ (see Appendix~\ref{appendix:FS_restriction} for  quantitative comparison).  Note, however, that the  frequency dependence  of the actual  $Z_n^{(0)}(\bar\mu)-1$ matches well the one of  $Z_n^{(0)}(\bar\mu)-1$  in FS-restricted  calculation. The same behavior holds for $Z_n^{(0)}(\bar\epsilon)$ at other $\bar\epsilon$.

We have also studied the additional component of the normal state self-energy, $\chi_n^{(0)}(\bar\mu)$ [Eq.~\eqref{eq:chi0}]. We find that for all values of $\bar\mu$,  $\chi_n^{(0)}(\bar\mu)$ is a weakly varying function of the Matsubara frequency, at least  for $\bar\omega_n<5$ which give the largest contribution to $\bar T_c$ (for details see Appendix~\ref{appendix:FS_restriction}).
 It is then safe to approximate $\chi_n^{(0)} (\epsilon)$ by a constant and absorb it  into the chemical potential $\bar\mu$. We assume that the same holds when we use the full Green's functions instead of the bare ones.

Neglecting $\bar\chi_n (\bar\epsilon)$, we reduce the full set of self-consistent  equations for the self-energy to coupled equations for the quasiparticle residue $Z_n (\bar\epsilon)$ and the pairing vertex $\phi_n (\bar\epsilon)$:
\begin{widetext}
\begin{align}
\label{eq:Zgapeq}
Z_{n}(\bar\epsilon)-1&=\frac{\bar T}{\bar\omega_n}\sum_{n'}u_{n-n'}
\int_0^{\bar\Lambda} d\bar\epsilon'\frac{N(\bar\epsilon')v(\bar\epsilon,\bar\epsilon')}{\left[\bar\omega_{n'} Z_{n'}(\bar\epsilon')\right]^2+\left[\bar\epsilon'-\bar\mu\right]^2}Z_{n'}(\bar\epsilon')\\
\label{eq:gapeq}
\phi_{n}(\bar\epsilon)&=\bar T\sum_{n'}u_{n-n'}\int_0^{\bar\Lambda} d\bar\epsilon'\frac{N(\bar\epsilon')v(\bar\epsilon,\bar\epsilon')}{\left[\bar\omega_{n'} Z_{n'}(\bar\epsilon')\right]^2+\left[\bar\epsilon'-\bar\mu\right]^2}\phi_{n'}(\bar\epsilon').
\end{align}
\end{widetext}

\begin{figure}
 \includegraphics[width=\columnwidth]{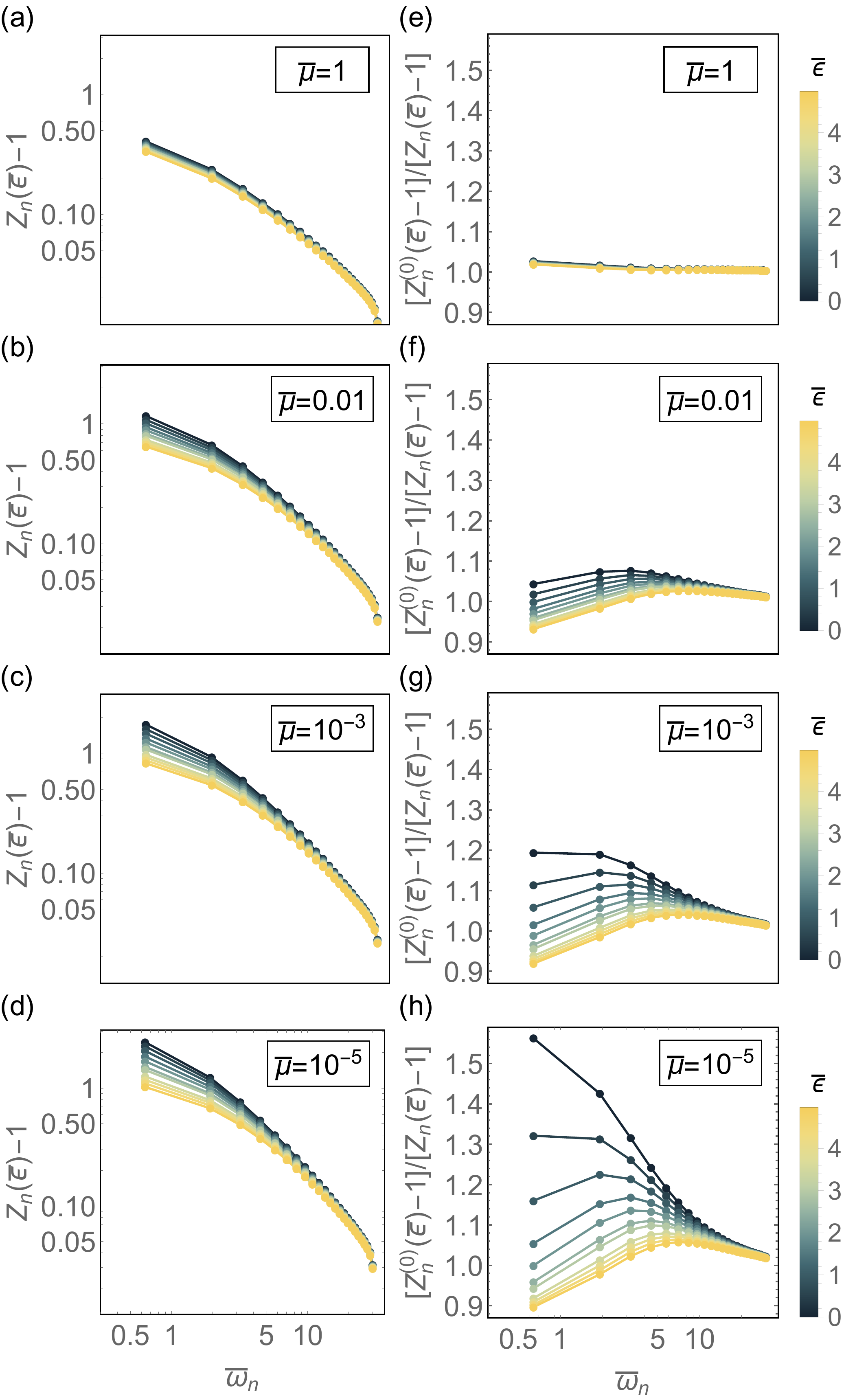}
\caption{\emph{The role of self-consistency  for  fermionic $Z_n(\bar\epsilon)$.}
Panels (a)-(d) --  $Z_n(\bar\epsilon)-1$  from Eq.~\eqref{eq:Zgapeq} vs $\bar\omega_n$ in a log-log scale, at various ${\bar \mu}$, indicated in legends.
 (e)-(h) The ratio between $Z^{(0)}_n(\bar\epsilon)-1$ computed with free fermions and the self-consistent expression $Z_n(\bar\epsilon)-1$ for those same $\bar\mu$.
}
\label{fig:NS_selfcons}
\end{figure}

In Fig.~\ref{fig:NS_selfcons} we show $Z_n (\bar \epsilon)$, obtained by the numerical solution of (\ref{eq:Zgapeq}), and compare it with $Z_n^{(0)} (\bar\epsilon)$.  We see that when $\bar\mu$ gets smaller,   the amplitude of $Z_n (\bar \epsilon)$  increases at all $\bar\epsilon$.
This is  consistent with the trend we found in Fig.~\ref{fig:NS_fsrestriction}(b), and we attribute it to the energy dependence of the coupling constant. In addition, we see from Fig.~\ref{fig:NS_selfcons}(b)-(d) that $Z_n (\bar \epsilon)$ acquires  significant energy dependence when  $\bar\mu$ decreases,   most notably for the first few Matsubara frequencies. These two features are concomitant by substantial corrections introduced by using the full fermionic propagators. We see from  Fig. \ref{fig:NS_selfcons}(e)-(h) that the stronger the amplitude and energy dependence of $Z_n (\bar \epsilon)$, the stronger the difference between the actual  $Z_n(\bar\epsilon)$, which comes out of a self-consistent calculation, and  $Z_n^{(0)}(\bar\epsilon)$ obtained using free fermions.

To summarize: at small $\bar\mu$ the contribution from electronic states away from the FS becomes dominant, and consequently, the integration over $\bar\epsilon'$ cannot be restricted to the vicinity of the FS.  This is the key new feature that invalidates the ME approximation. The modification introduced by self-consistency on $Z_n (\bar\epsilon)$, on the other hand, is rather modest and not fundamentally relevant. We expect that vertex corrections (the corrections to ladder approximation) will also be at most modest.
 We now proceed to the calculation of  superconducting $\bar T_c$.

\subsection{Critical temperature beyond ME}

\begin{figure}
 \includegraphics[width=\columnwidth]{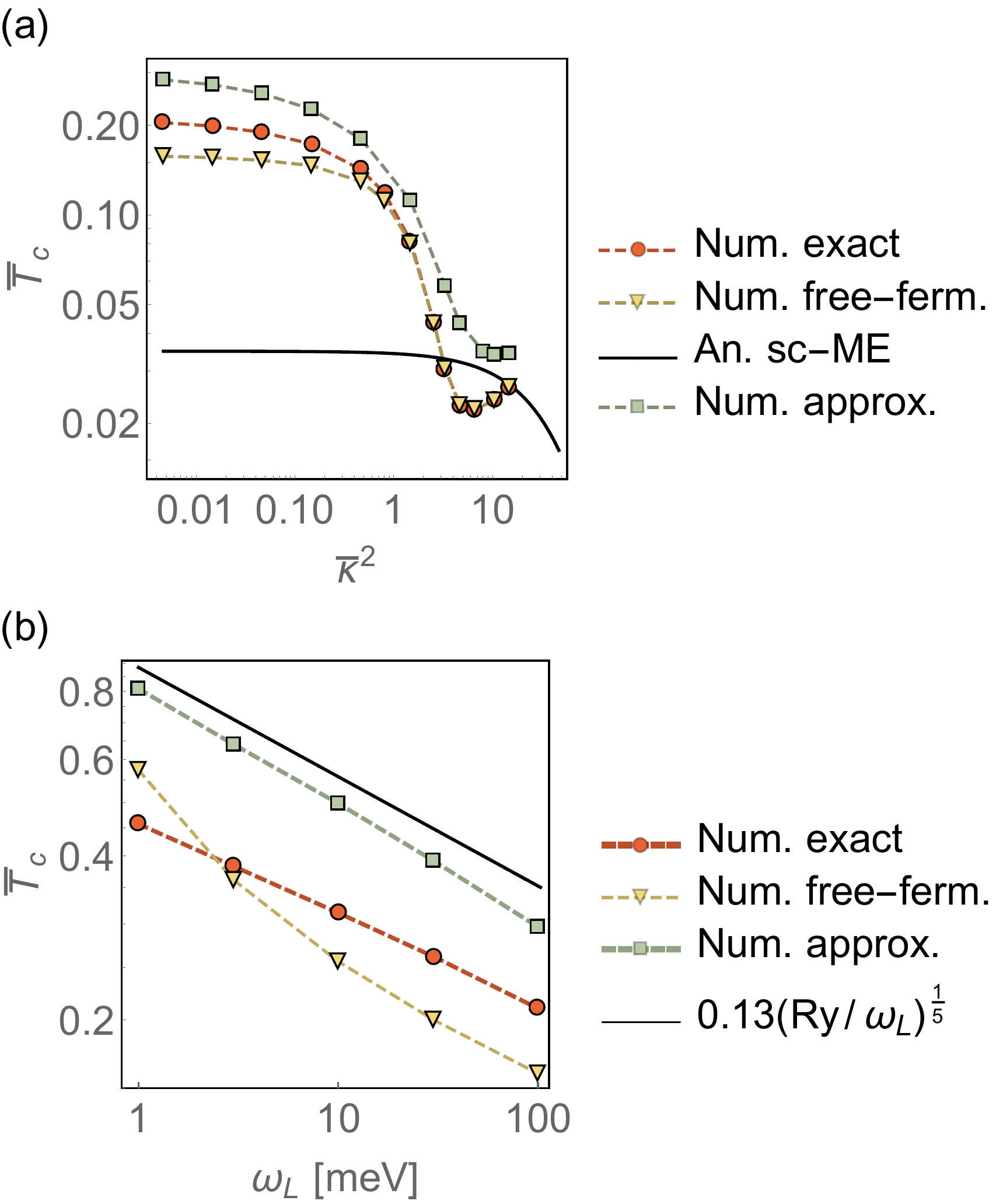}
\caption{\emph{Superconducting
$\bar T_c=\frac{T_c}{\omega_L}$
}.
(a) $\bar T_c$ versus the rescaled screening $\bar\kappa^2=\frac{\kappa^2/2m}{\omega_L}=\frac{4}{\pi}\sqrt{\bar\mu\bar\rho}$.
 Red circles -- the numerical solution of Eq.~\eqref{eq:gapeq} with the actual $Z_n (\bar \epsilon)$  given by Eq.~\eqref{eq:Zgapeq};
  yellow triangles -- the numerical solution of Eq.~\eqref{eq:gapeq} with the approximate (non-self-consistent) $Z_n^{(0)} (\bar \epsilon)$, computed with free fermions; black line -- the ME expression at small coupling, Eq.~\eqref{eq:tc_elias}; green squares -- the
  numerical solution of the approximate gap equation~\eqref{eq:gapeq_neqn}.
      In all cases we set $\omega_L=100$ meV.
(b) Numerical solutions in the $\bar\mu=0$ limit as a function of phonon frequency $\omega_L$ for three
  computational procedures described for panel (a) (specified in the legend). The black line is the analytical expression $\bar T_c^{\text{an}}$ [Eq.~\eqref{eq:tc_mu0_approx}].}
\label{fig:pairing_off_fs}
\end{figure}

We solved the set of linearized gap equations~\eqref{eq:Zgapeq}-\eqref{eq:gapeq} numerically for different ${\bar \mu}$. The results are presented in Fig. \ref{fig:pairing_off_fs}(a) (red circles).  For comparison, we  also present the results for $\bar T_c$ using $Z_n^{(0)} (\bar \epsilon)$ instead of the actual $Z_n (\bar \epsilon)$ (yellow down-triangles) and the  analytical expression,  $\bar T_c^{\text{An.sc-ME}} (\bar \mu)$, Eq.~\eqref{eq:tc_elias}, obtained within the ME formalism at small coupling (black curve). We see that at small $\bar \mu$, the actual
$\bar T_c (\bar \mu)$ is substantially larger than the approximate small-coupling ME result $\bar T_cs$. To be more specific, the actual $\bar T_c (\bar \mu)$ saturates  at around $\bar T_c\sim 0.15$ in the limit $\bar\mu\rightarrow 0$, while $\bar T_c^{\text{An.sc-ME}}\sim 0.034$ in this limit.  The difference between the actual $\bar T_c$ and the one obtained by evaluating $Z$ with free fermions is about  $30\%$  at small ${\bar \mu}$.
At large $\bar \mu$, ${\bar T}_c (\bar \mu)$ approaches $\bar T_c^{\text{An.sc-ME}} (\bar \mu)$, as expected.

To understand why $\bar T_c$ is quite high at $\bar\mu \rightarrow 0$  and what is its dependence on $\omega_L$ in this limit, we now obtain an approximate analytical solution of Eqs.~\eqref{eq:Zgapeq}-\eqref{eq:gapeq}.
We first note that, when $Z$ is large, like, e.g., when the system is near a quantum-critical point, the key effect of $Z$ is to cancel the potentially singular  $n = n'$ term in the gap equation (such cancellation can be seen in  Eq. \eqref{eq:gapeq_delta}); other effects of $Z$ are not important.   Accordingly, we restrict the frequency summation in the gap equation~\eqref{eq:gapeq} to $n'\neq n$ and, after that, set $Z_{n'}(\bar\epsilon')=1$.
This simplifies the gap equation to
\begin{align}
\label{eq:gapeq_neqn}
  \phi_n(\bar\epsilon)&=\bar T\sum_{n'\neq n}u_{n-n'}
\int_0^{\bar\Lambda} d\bar\epsilon'\frac{N(\bar\epsilon')v(\bar\epsilon,\bar\epsilon')}{\bar\omega_{n'}^{2}+(\bar\epsilon'-\bar\mu)^2}
 \phi_{n'}(\bar\epsilon').
\end{align}
In order to gauge the accuracy of reducing the effect of $Z_n(\bar\epsilon)$ to the cancellation of $n=n'$ terms in the equation for the pairing vertex,
we solved Eq. \eqref{eq:gapeq_neqn} for  $\bar T_c$ numerically. We plot  the result as green squares in Fig.~\ref{fig:pairing_off_fs}(a) and label it as an "approximate" numerical solution.  The evolution of this $\bar T_c (\bar \mu)$  with $\bar\mu$ nearly matches that of the actual $\bar T_c$ (red circles), while the magnitude is larger by about $35\%$.  This is satisfactory for our purposes, as our goal  is to understand the evolution of $\bar T_c (\bar \mu)$ with $\bar \mu$ and which combination of parameters sets the scale for ${\bar T}_c (\bar \mu)$ at small $\bar \mu$.

Having verified the validity of  Eq. \eqref{eq:gapeq_neqn}, we continue with the analytic analysis. We conjecture that in the $\bar\mu=0$ limit, this equation can be simplified further. Based again on potential analogy with the pairing problem near a Quantum-Critical point, we assume and then verify that to extract  ${\bar T}^{\text{an}}_c (\bar \mu=0)$ from  \eqref{eq:gapeq_neqn} one can restrict the sum to
the first few Matsubara frequencies, more specifically to $\bar\omega_n = \pm \pi \bar T$.  This approximation eliminates the frequency summation in   \eqref{eq:gapeq_neqn} and reduces it to a 1D integral equation for $\phi_0 (\bar\epsilon) = \phi_{-1} (\bar\epsilon)$:
\begin{align}
\label{eq:gapeq_neqn_n0}
  {\tilde \phi}_0(\bar p)& = \frac{2\sqrt{\bar\rho}\bar T/\pi}{1+(2\pi \bar T)^2}\int_0^{\infty} d\bar p'\frac{\log \left(\frac{\bar p+ \bar p'}{|\bar p-\bar p'|}\right)}{(\pi\bar T)^{'2}+\bar p^{'4}}{\tilde \phi}_0(\bar p').
\end{align}
Here we switched to the radial momentum variable $\bar p'=\sqrt{\bar\epsilon'}$, and introduced the rescaled variable
${\tilde \phi}_0 (\bar p) = \bar p\phi_0(\bar p)$.
Solving this equation (see Appendix~\ref{appendix:tcmu0} for details), we obtain the following relation for $\bar T_c$,
\begin{equation}
\label{eq:tc_mu0_full}
 \sqrt{\pi {\bar T}^{\text{an}}_c}\left[1+4(\pi \bar T^{\text{an}}_c)^2\right]=\frac{8}{(5-\sqrt{10})\pi^2}\sqrt{\bar\rho},
\end{equation}
where, we remind $\bar\rho=\frac{\text{Ry}}{\omega_L}$.
Because the Rydberg energy ($\text{Ry}=13.6$ eV) is in general much larger than the characteristic phonon frequency $\omega_L$, i.e. $\bar\rho\gg 1$, Eq. (\ref{eq:tc_mu0_full}) can be solved by expanding in $1/{\bar \rho}$.  We then obtain
\begin{equation}
\label{eq:tc_mu0_approx}
 {\bar T}^{\text{an}}_c\simeq\frac{1}{\pi}\left[\frac{2}{\pi^2(5-\sqrt{10})}\right]^{\frac{2}{5}}\bar\rho^\frac{1}{5}
 \longrightarrow  T^{\text{an}}_c\simeq 0.13\omega_L\left(\frac{\text{Ry}}{\omega_L}\right)^{\frac{1}{5}}
\end{equation}
 We see that, in the actual dimension-full units,  $ T^{\text{an}}_c$ contains $\omega_L$ as the overall factor, but also contains an enhancement factor $\text{Ry}/\omega_L$ with the non-trivial exponent $1/5$.

We verified this functional form by plotting in Fig.~\ref{fig:pairing_off_fs}(b) various numerical ${\bar T}_c$'s  from  Fig.~\ref{fig:pairing_off_fs}(a) as functions of $\omega_L$.  We see that the actual ${\bar T}_c$, and ${\bar T}_c$ obtained in  approximate  numerical calculations follow $\omega^{-1/5}_L$ behavior quite well.  As an independent check, we fitted the actual  ${\bar T}_c$ by $\omega^{-\eta}_L$ and obtained $\eta \approx 1/6$, which is  quite close to the analytical result $\eta = \frac{1}{5}$.

\section{Summary}
\label{section:summary}

Phonon-mediated superconductivity has been extensively studied within the Migdal-Eliashberg framework, which is suitable for conventional metals with $\mu\gg\omega_L$.
In this work, we analyzed pairing in the opposite regime of low carrier density, where $\mu\ll\omega_L$, which is believed to be relevant to several bulk systems, most notably SrTiO$_3$ and Bi. By considering the attractive part of the Coulomb-screened electron-phonon interaction, both the frequency and momentum dependencies of the pairing interaction were included.
We showed that the contribution to the mass renormalization function and pairing vertex are dominated by electronic states away from the FS. This result is in stark contrast to the more standard regime $\mu\gg\omega_L$, where these processes can be safely neglected and the computation of the fermionic self-energy can be confined to the vicinity of the FS. More specifically, we solved the momentum- and frequency-dependent gap equations to obtain the pairing instability temperature $T_c$ as a function of $\mu/\omega_L$.
We found, both numerically and analytically, a substantial increase of the critical temperature $T_c$ in the limit $\mu\rightarrow 0$.
In fact, the value of $T_c$ not only remains finite in this limit, but exceeds the characteristic phonon frequency  $\omega_L$.
We obtained the analytical result
   $T_c\propto\omega_L \left(\frac{\text{Ry}}{\omega_L}\right)^{\eta}$,  with $\eta=\frac{1}{5}$, for $\mu\rightarrow 0$, which agrees well with our numerical results for $T_c(\omega_L)$.
Future investigations that include also the repulsive part of the Coulomb-screened electron-phonon interaction are desirable, particularly since in the $\mu\ll\omega_L$ limit there is a very narrow window for the renormalization of the Coulomb repulsion into a reduced effective pseudopotential.\\

\begin{acknowledgments}
We thank K. Behnia, G. Lonzarich, D.L. Maslov and A. Aperis for fruitful discussions. This work was supported by the U. S. Department of Energy through the University of Minnesota Center for Quantum Materials, under Award No. DE-SC-0016371.
\end{acknowledgments}

\appendix
\section{Effective interaction and the $\omega_q\rightarrow\omega_L$ approximation}
\label{appendix:interaction}
The difference between the renormalized phonon frequency $\omega_q$ and the bare frequency $\omega_L$ has not been included in the electron-phonon interaction Eq.~\eqref{eq:Veff}. Here we show that the renormalization of $\omega_L$ can be neglected for the superconducting channel within the leading logarithmic approximation. The term that has not been included in Eq.~\eqref{eq:Veff} is
\begin{equation}
\label{eq:fnq}
f_n(q)=-\frac{\bar \Omega_n^2}{1+\bar \Omega_n^2}\frac{\kappa^2}{q^2\left(1+\bar \Omega_n^2\right)+\kappa^2\bar \Omega_n^2}.
\end{equation}
We follow the notation in the main text, with an overbar denoting a rescaled variable with respect to the bare phonon frequency, $\bar\Omega_n=\frac{\Omega_n}{\omega_L}$. After the angular integration is performed, the s-wave component of the electron-phonon term when $\omega_q$ is included is
\begin{equation}
\label{eq:Velph_full}
V^{\text{el-ph}}_n(\bar p,\bar p')=\frac{1}{1+\bar\Omega_n^2}\left[v(\bar p,\bar p')+\mathcal{F}_n(\bar p,\bar p')\right],
\end{equation}
where $v(\bar p,\bar p')$ is the radial momentum dependent term that has been included in our model [Eq.~\eqref{eq:vpp}], and $\mathcal{F}_n(\bar p,\bar p')$ is the additional contribution coming from the renormalized frequency, the $f_n(q)$ term [Eq.~\eqref{eq:fnq}]:
\begin{widetext}
\begin{equation}
\mathcal{F}_n(\bar p,\bar p')=-\bar\Omega_n^2 \frac{\sqrt{\bar\rho}}{2\pi\bar p\bar p'}\log\left[1+\frac{4\bar\kappa^2\bar p\bar p'}{\left[(\bar p-\bar p')^2(1+\bar\Omega_n^2)+\bar\Omega_n^2\bar\kappa^2\right]\left[(\bar p+\bar p')^2+\bar\kappa^2\right]}\right].
\end{equation}
\end{widetext}
We now insert the electron-phonon interaction \eqref{eq:Velph_full} in the fermionic self-energy equations \eqref{eq:Zfull}-\eqref{eq:phifull} and apply the ME approximations described in the main text. By expanding for small coupling to first order in $\lambda$, $Z_n=1+\lambda+\mathcal{O}(\lambda^2)$ and we get for the pairing vertex the following equation,
\begin{equation}
\phi_n=\lambda^* \sum_{n'}\chi_{n-n'}\frac{\pi \bar T}{|\bar\omega_{n'}|}\phi_{n'}
\end{equation}
instead of Eq.~\eqref{eq:gapeq_fs_*}, where the propagator $\chi_{n-n'}$ has acquired an additional term,
\begin{widetext}
\begin{equation}
\chi_{n-n'}=\frac{1}{1+|\bar\omega_n-\bar\omega_{n'}|^2}-\frac{|\bar\omega_n-\bar\omega_{n'}|^2}{1+|\bar\omega_n-\bar\omega_{n'}|^2}\frac{\log\left(1+\frac{1}{|\bar\omega_n-\bar\omega_{n'}|^2}\frac{1}{1+\alpha}\right)}{\log\left(1+\frac{1}{\alpha}\right)}
\end{equation}
\end{widetext}
with $\alpha=\frac{1}{\pi}\sqrt{\frac{\text{Ry}}{\mu}}$. In order to get an estimate of the contribution of this additional term in the propagator to $T_c$, we neglect the $\bar\omega_n$ dependence in $\chi_{n-n'}$, and the gap equation is further simplified
\begin{align}
1&=\lambda^*\sum_{n'}\frac{1}{1+|\bar\omega_{n'}|^2}\left[1-|\bar\omega_{n'}|^2\frac{\log\left(1+\frac{1}{|\bar\omega_{n'}|^2}\frac{1}{1+\alpha}\right)}{\log\left(1+\frac{1}{\alpha}\right)}\right]\frac{\pi \bar T}{|\bar\omega_{n'}|}\\\nonumber
\quad&\simeq -\lambda^*\log(\pi\bar T)+\mathcal{O}(\lambda^*).
\end{align}
As seen, the leading logarithmic contribution is given by the first term (which we have included in our model), and the contribution from the second term (coming from the renormalization of $\omega_L$ by the electronic screening) is linear in $\lambda^*$ and without a logarithm. We thus neglect this contribution and use the effective interaction~\eqref{eq:Veff}.

\section{The derivation of the gap equations}
\label{appendix:eqsderivation}

In this section we present the details of the derivation of the fermionic self-energy equations~\eqref{eq:Zfull}-\eqref{eq:phifull} for the effective electron-phonon interaction,
\begin{equation}
 V_{n}^{\text{e-ph}}(q)=-\frac{4\pi e^2}{q^2+\kappa^2}\frac{\omega_L^2}{\Omega_n^2+\omega_L^2}.
\end{equation}
The various components of the equations are in terms of the dimension-full variables,
\begin{widetext}
\begin{align}
\frac{1}{\omega_L}\int_0^{\sqrt{2m\Lambda}}\frac{dk'k^{'2}}{2\pi}\frac{1}{2\pi}\int_{-1}^1dx P_{l=0}(x)\frac{4\pi e^2}{k^2+k^{'2}-2kk'x+\kappa^2}&=\int_0^{\bar\Lambda}d\bar\epsilon'N(\bar\epsilon')v(\bar\epsilon,\bar\epsilon') \\
\frac{\omega_L^2}{\omega_L}\frac{T}{\left[\omega_{n'}Z_{n'}(k')\right]^2+\left[\epsilon'-\mu+\chi_{n'}(k')\right]^2}&=\frac{\bar T}{\left[\bar\omega_{n'}Z_{n'}(\bar\epsilon')\right]^2+\left[\bar\epsilon'-\bar\mu+\bar\chi_{n'}(\bar\epsilon')\right]^2},
\end{align}
\end{widetext}
with the first Legendre polynomial $P_{l=0}(x)=1$ for the isotropic s-wave component of the interaction.

\section{FS restriction of $Z_n^{(0)}(\bar\mu)$ and $\bar\chi_n^{(0)}(\bar\mu)$}
\label{appendix:FS_restriction}

\begin{figure*}
 \includegraphics[width=0.9\textwidth]{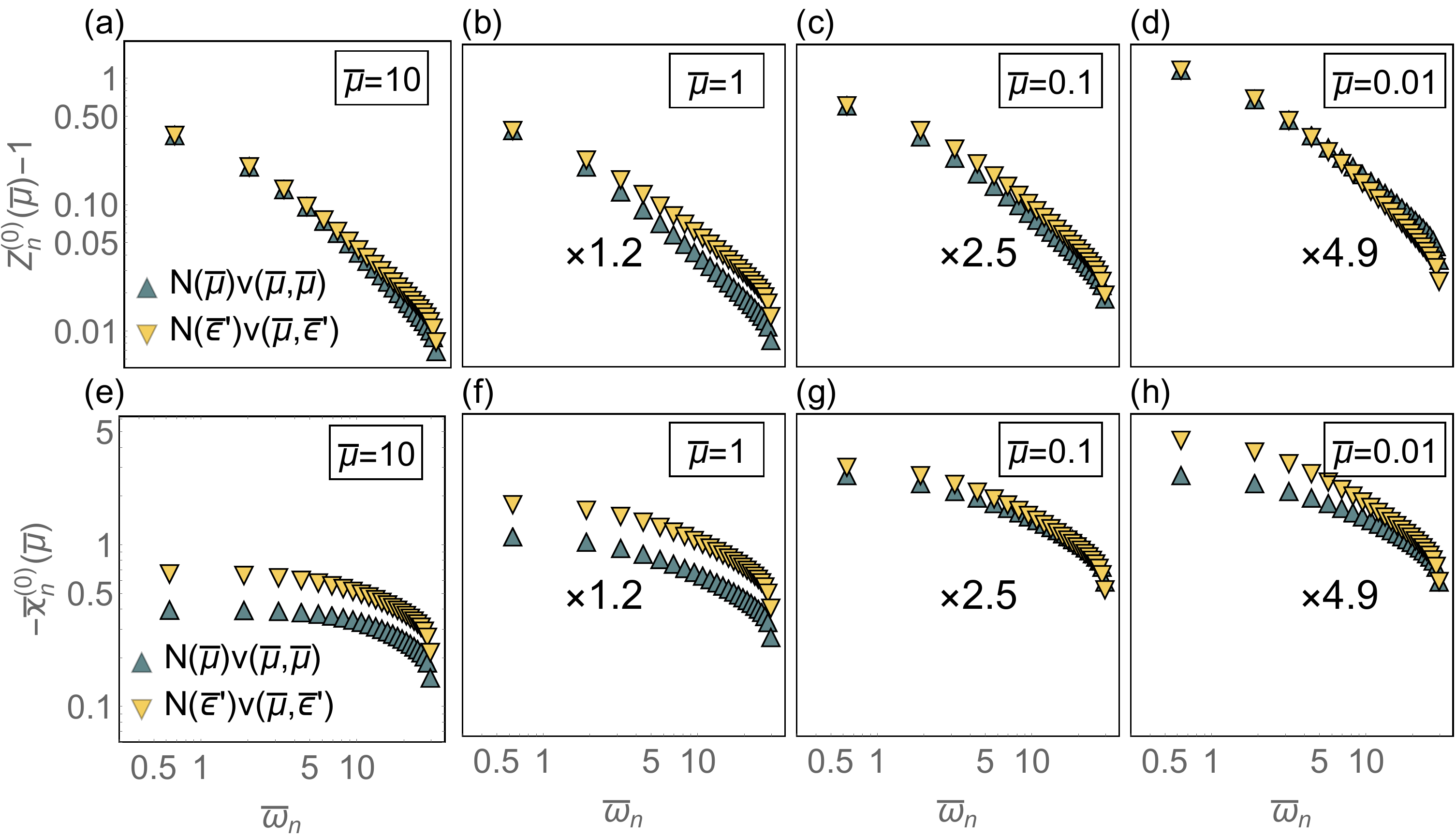}
\caption{Mass renormalization $Z_n^{(0)}(\bar\mu)-1$ vs $\bar\omega_n$ in a log-log scale for (a) $\bar\mu=10$, (b) $\bar\mu=1$, (c) $\bar\mu=0.1$, and (d) $\bar\mu=0.01$ computed with FS restriction (blue up-triangles) and keeping the full energy dependence (yellow down-triangles). The blue data has been renormalized by the number specified in each panel to match the first Matsubara frequency of the yellow data. Panels (e)-(h) -- the same  plots for the energy-shift function $-\bar\chi_n^{(0)}(\bar\mu)$.
In all panels $\bar T=0.2$.}
\label{fig:chi0}
\end{figure*}

Figure~\ref{fig:chi0} shows the normal state components of the self-energy $Z_n^{(0)}(\bar\mu)-1$ and $\bar\chi_n^{(0)}(\bar\mu)$, which were discussed in Section~\ref{subsection:validity}. They are computed in two different ways: (i) by keeping the energy dependence of the coupling constant $N(\bar\epsilon')v(\bar\epsilon,\bar\epsilon')$ and (ii) restricting the calculation to the FS vicinity $N(\bar\mu)v(\bar\mu,\bar\mu)$. Panels (a)-(d) illustrate the quantitative comparison of the mass renormalization computed with and without the FS restriction (blue and yellow data, respectively). The FS restricted function has been renormalized by the number specified in each panel to match the $n=0$ Matsubara frequency of the non-FS restricted function. For high $\bar\mu$, $Z_n^{(0)}(\bar\mu)-1$ computed with and without the FS restriction are almost identical. As the chemical potential is reduced, the non-FS restricted computation (yellow data) results in higher mass renormalization. On the contrary, the FS restricted result (blue data) stays nearly unchanged and for a quantitative comparison, we show this data renormalized by the number specified in each panel [Figs.~\ref{fig:chi0}(b)-(d)]. It reaches almost a factor of $5$ difference for $\bar\mu=0.01$, which highlights the dominant contribution of the states far from the FS in the $\bar\mu\ll 1$ regime.

The energy shift function $-\bar\chi_n^{(0)}(\bar\mu)$ is also shown for the same parameters in Fig.~\ref{fig:chi0}(e)-(h). As seen, the electronic processes away from the FS dominate this self-energy component as well in the dilute $\bar\mu\ll 1$ limit. The FS-restricted results have been renormalized by the same number as in panels (a)-(d) for clarity. In all cases $\bar\chi_n^{(0)}(\bar\mu)$ is a weakly varying function of $\bar\omega_n$ for $\bar\omega_n<5$.

\section{Derivation of the $\bar T_c$ expression at $\bar\mu=0$}
\label{appendix:tcmu0}

We give below the detailed solution of the gap equation~\eqref{eq:gapeq_neqn_n0},
 \begin{equation}
  \phi(\bar p)=\frac{2\sqrt{\bar\rho}\bar T/\pi}{1+(2\pi \bar T)^2}\int_0^{\infty} d\bar p'\frac{\log \left(\frac{\bar p+ \bar p'}{|\bar p-\bar p'|}\right)}{(\pi\bar T)^{'2}+\bar p^{'4}}\phi(\bar p')
\end{equation}
 to obtain the relation for $\bar T_c$ given by Eq.~\eqref{eq:tc_mu0_full}.
By making the change of variables $x=\frac{\bar p'}{\sqrt{\pi \bar T}}$ and $y=\frac{\bar p}{\sqrt{\pi \bar T}}$ the equation becomes
\begin{equation}
  \phi(y)=C \int_0^{\infty} dx\frac{1}{1+x^{4}}\log \left(\frac{y+x}{|y-x|}\right)
 \phi(x)
\end{equation}
where
\begin{equation}
\label{eq:C_coeff}
 C=\frac{2\sqrt{\rho}}{\pi^2}\frac{1}{\sqrt{\pi \bar T}\left[1+(2\pi \bar T)^2\right]}
\end{equation}
and we have taken $\sqrt{\bar\Lambda/\pi \bar T}\rightarrow \infty$.
We then consider the two limits $x\gg y$ and $x\ll y$ and approximate the logarithm and gap function accordingly,
\begin{equation}
 \log \left(\frac{1+z}{|1-z|}\right)=
\begin{cases}
  2/z & \mbox{if } z\gg 1\\
  2z & \mbox{if } z\ll 1
\end{cases}
\end{equation}
with $z=\frac{x}{y}$.
We then look for gap function solutions of the form
\begin{equation}
\phi(z)=
\begin{cases}
 A/z & \mbox{if } z\gg 1\\
 Bz & \mbox{if } z\ll 1
\end{cases}
\end{equation}
and the get the following set of coupled equations,
\begin{equation}
 \begin{cases}
  \frac{A}{y}=C \left(\int_0^1 dx \frac{2x}{y} Bx + \int_1^{\infty}dx \frac{1}{x^4}\frac{2x}{y} \frac{A}{x}\right) \\
  By=C\left(\int_0^1dx\frac{2y}{x} Bx + \int_1^{\infty}dx \frac{1}{x^4}\frac{2y}{x}\frac{A}{x}\right).
 \end{cases}
\end{equation}
Solving these equations we get the condition for $C$
\begin{equation}
\begin{vmatrix}
 1-\frac{2}{3}C & \frac{2}{3}C\\
 \frac{2}{5}C & 2C-1
 \end{vmatrix}=0,
\end{equation}
which gives $C=\frac{1}{4}\left(5-\sqrt{10}\right)$ at $\bar T_c$. Finally, by equating this coefficient to the initial temperature dependent expression in Eq.~\eqref{eq:C_coeff} we recover Eq.~\eqref{eq:tc_mu0_full} for $\bar T_c$ at $\bar\mu=0$,
\begin{equation}
 \frac{2\sqrt{\bar\rho}}{\pi^2}\frac{1}{\sqrt{\pi \bar T_c}\left[1+(2\pi \bar T_c)^2\right]}=\frac{1}{4}\left(5-\sqrt{10}\right).
\end{equation}

\bibliography{bibliography}

\end{document}